\documentstyle[twoside,fleqn,espcrc2,%
times,oldlfont,amssymb,amsbsy,epsfig]{article}

\title{A Model Independent Analysis of Solar Neutrino Data}

\author{S.M. Bilenky%
\address{%
INFN Torino,
and
Dip. di Fisica Teorica, Universit\`a di Torino,
Via P. Giuria 1, 10125 Torino, Italy
}%
\address{%
Joint Institute for Nuclear Research, Dubna, Russia
}%
\thanks{E-mail address: BILENKY@TO.INFN.IT.}%
and
C. Giunti$^{\mathrm{a}}$%
\thanks{E-mail address: GIUNTI@TO.INFN.IT.}%
}

\begin{document}
\pagestyle{empty}

\begin{abstract}
We present the results of a solar model independent analysis
of the existing solar neutrino data
in the case of
$\nu_{e}\to\nu_{\mu(\tau)}$
MSW transitions.
We considered two cases.
In the first case
no assumptions on the values of the total fluxes of neutrinos
from all reactions have been made.
In the second case
we constrained
the values of the neutrino fluxes
within some wide limits
that take into account the predictions
of the existing Standard Solar Models.
We show that in both cases
the existing data allow to exclude
rather large regions
in the plane of the parameters
$ \Delta m^2 $ and $ \sin^2 2\theta $.
\end{abstract}

\maketitle
\setcounter{footnote}{0}

The problem of neutrino
masses and mixing
is one of the most significant problems
of modern physics.
The investigation of this problem
is connected with the
possibility to discover
some effects of new physics
beyond the Standard Model.
Solar neutrino experiments
are very important for this investigation.
These experiments are unique
because they
are sensitive to very small values
of the difference of the squared neutrino masses
(down to
$ \Delta m^2 \simeq \times 10^{-10} \, \mathrm{eV}^2 $)
and
to a wide region of mixing angles
$\theta$,
including very small ones.

At present there exist data of four solar neutrino experiments:
the radiochemical experiments
Homestake
\cite{B:HOMESTAKE},
GALLEX
\cite{B:GALLEX}
and SAGE
\cite{B:SAGE}
and the direct counting experiment
Kamiokande~\cite{B:KAMIOKANDE}.
The event rates observed in all these four experiments
are much smaller
than the rates predicted by the
Standard Solar Model (SSM)
\cite{B:BAHCALL,B:SACLAY,B:CDF,B:SBF,B:NICE}.
This ``leakage''
of solar neutrinos can be very naturally
explained by neutrino mixing
(see Ref.\cite{B:BILENKY})
and resonant matter effects (MSW)
\cite{B:MSW}.
In fact,
all the existing data
can be described
by the MSW mechanism
in the simplest case of mixing
between two neutrino types.
Two MSW solutions have been found
\cite{B:AR}:
a small mixing angle solution with
$ \Delta m^2 \simeq 5 \times 10^{-6} \, \mathrm{eV}^2 $
and
$ \sin^2 2\theta \simeq 8 \times 10^{-3} $
and
a large mixing angle solution with
$ \Delta m^2 \simeq 10^{-5} \, \mathrm{eV}^2 $
and
$ \sin^2 2\theta \simeq 0.8 $.
However,
let us emphasize that
these solutions are based on the assumption
that the neutrino fluxes
from the different reactions
are given by the SSM.
As it is well known,
the neutrino fluxes predicted by the SSM are subject
to many sources of uncertainties,
mainly due to a poor knowledge of some input parameters
(especially nuclear cross sections and solar opacities).
Taking into account the importance of neutrino mixing
for the physics beyond the Standard Model,
any model independent information
on neutrino mixing
(i.e. independent from the SSM)
that can be obtained from the existing data
is of great interest.
In this report we present the results
of such a model independent analysis
in the case of mixing between two active neutrino states
($\nu_{e}$--$\nu_{\mu}$ or $\nu_{e}$--$\nu_{\tau}$)\footnote
{
The case of mixing between $\nu_e$
and sterile neutrinos
was considered in Ref.\cite{B:BG94B}
}.
We consider two cases.
In the first case we do not make any
assumption on the possible values
of the solar neutrino fluxes.
In the second case
we constrain the values
of the solar neutrino fluxes
within some wide limits
that take into account
the predictions of the
Standard Solar Models
\cite{B:BAHCALL,B:SACLAY,B:CDF,B:SBF,B:NICE}.
The existing solar neutrino data
are not sufficient to determine
allowed regions
for the parameters
$ \Delta m^2 $ and $ \sin^2 2\theta $
in the two cases
under consideration.
Instead,
we will show that
the existing solar neutrino data
allow to {\em exclude}
rather large regions
of the parameters
$ \Delta m^2 $ and $ \sin^2 2\theta $,
even in the first case
in which no restriction
is imposed on the possible
values of the solar neutrino fluxes.

Let us write the initial
spectrum of $\nu_{e}$
from the source $r$
($r$ $=$
$pp$, $pep$, $^7\mathrm{Be}$, $^8\mathrm{B}$, $\mathrm{He}p$,
$^{13}\mathrm{N}$, $^{15}\mathrm{O}$, $^{17}\mathrm{F}$)
in the form
\begin{equation}
\phi^{r}(E)
=
X^{r}(E)
\,
\Phi^{r}
\;,
\label{E1011}
\end{equation}
where
$E$ is the neutrino energy,
$ X^{r}(E) $
is a known normalized function
(see Ref.\cite{B:BAHCALL})
and
$ \Phi^{r} $
is the total initial neutrino flux.
The integral event rate
in any experiment $a$
($a$ = HOM (Homestake), GAL (GALLEX+SAGE), KAM (Kamiokande))
is given by the expression
\begin{equation}
N_{a}
=
\sum_{r}
Y_{a}^{r}
\,
\Phi^{r}
\;.
\label{E1001}
\end{equation}
In our calculations we used
$
N_{\mathrm{HOM}}^{\mathrm{exp}}
=
2.32 \pm 0.23 \, \mathrm{SNU} 
$
\cite{B:HOMESTAKE},
the combined
GALLEX--SAGE event rate
$
N_{\mathrm{GAL}}^{\mathrm{exp}}
=
74 \pm 9 \, \mathrm{SNU} $
\cite{B:GALLEX,B:SAGE}
and
$
N_{\mathrm{KAM}}^{\mathrm{exp}} / N_{\mathrm{KAM}}^{\mathrm{BP}}
=
0.51 \pm 0.04 \pm 0.06
$
\cite{B:KAMIOKANDE},
where
$ N_{\mathrm{KAM}}^{\mathrm{BP}} $
is the Kamiokande event rate predicted by BP
\protect\cite{B:BAHCALL}.

In the radiochemical experiments
only solar $\nu_{e}$ are detected.
We have
\begin{equation}
Y_{a}^{r}
=
\int_{E_{\mathrm{th}}^{a}}
\sigma_{a}(E)
\,
X^{r}(E)
\,
\mathrm{P}_{\nu_{e}\to\nu_{e}}(E)
\,
\mathrm{d} E
\;,
\label{E1004}
\end{equation}
with
$a=\mathrm{HOM},\mathrm{GAL}$.
Here
$ \sigma_{a}(E) $
is
the neutrino cross section,
$ \mathrm{P}_{\nu_{e}\to\nu_{e}}(E) $
is the probability of $\nu_{e}$ to survive
and
$ E_{\mathrm{th}}^{a} $
is the threshold energy.
For the calculation of
the $\nu_{e}$ survival probability
we used the formula given in Ref.\cite{B:PEEMSW},
which is valid for an exponentially decreasing
electron density.

In the the Kamiokande experiment
$\nu_{e}$ as well as
$\nu_{\mu}$
and
$\nu_{\tau}$
are detected.
We have
\begin{eqnarray} \nopagebreak
\null \hspace{-0.5cm}
&&
Y_{\mathrm{KAM}}^{r}
=
\int_{E_{\mathrm{th}}^{\mathrm{ES}}}
\sigma_{\nu_{e}e}(E)
\,
X^{r}(E)
\,
\mathrm{P}_{\nu_{e}\to\nu_{e}}(E)
\,
\mathrm{d} E
\nonumber
\\
\null \hspace{-0.5cm}
&&
+
\int_{E_{\mathrm{th}}^{\mathrm{ES}}}
\sigma_{\nu_{\mu}e}(E)
\,
X^{r}(E)
\,
\sum_{\ell=\mu,\tau} \mathrm{P}_{\nu_{e}\to\nu_{\ell}}(E)
\,
\mathrm{d} E
\;,
\label{E1003}
\end{eqnarray}
where
$ \sigma_{\nu_\ell e}(E) $
is the cross section of the process
$ \nu_\ell \, e \to \nu_\ell \, e $
($\ell=e,\mu$),
$ E_{\mathrm{th}}^{\mathrm{ES}} $
is the recoil electron energy threshold
and
$ \mathrm{P}_{\nu_{e}\to\nu_{\ell}}(E) $
is the probability of the transition
$ \nu_{e} \to \nu_{\ell} $
($ \ell = e , \mu $).
In our calculation we took into account
the efficiency and the energy resolution
of the Kamiokande detector
\cite{B:KAMIOKANDE}.

The fluxes of neutrinos produced
in the thermonuclear $pp$ and CNO
cycles must satisfy the following constraint:
\begin{equation}
N_{\mathrm{LUM}}
=
\sum_{r}
Y_{\mathrm{LUM}}^{r}
\,
\Phi^{r}
\;.
\label{E1012}
\end{equation}
Here
$ \displaystyle
N_{\mathrm{LUM}}
=
L_{\odot}
/
4 \pi \, d^2
=
( 8.491 \pm 0.018 ) \times 10^{11}
\, \mathrm{MeV} \, \mathrm{cm}^{-2} \, \mathrm{sec}^{-1}
$,
where
$ d = 1 \, \mathrm{AU} $
is the average sun--earth distance,
$ L_{\odot} $
is the luminosity of the sun.
The factors
$ \displaystyle
Y_{\mathrm{LUM}}^{r}
=
Q / 2 - \left\langle E \right\rangle^{r}
$
(where
$ Q = 4 \, m_{p} + 2 \, m_{e} - m_{^4\mathrm{He}}
    = 26.73 \, \mathrm{MeV} $
and
$ \left\langle E \right\rangle^{r} $
is the average energy of neutrinos from the source $r$)
are given in Table \ref{T:FLUXES}.

\begin{table}[t]
\begin{tabular*}{\linewidth}
{c@{\extracolsep{\fill}}
 c@{\extracolsep{\fill}}
 c@{\extracolsep{\fill}}
 c}
\hline
\hline
Source
&
$ Y_{\mathrm{LUM}} (\mathrm{MeV}) $
&
$ \xi_{\mathrm{min}} $
&
$ \xi_{\mathrm{max}} $
\\
\hline
$pp$
&
13.10
&
0.93
&
1.07
\\
$pep$
&
11.92
&
0.61
&
1.29
\\
$^7\mathrm{Be}$
&
12.55
&
0.46
&
1.40
\\
$^8\mathrm{B}$
&
6.66
&
0
&
1.43
\\
$\mathrm{He}p$
&
3.74
&
0.90
&
1.13
\\
$^{13}\mathrm{N}$
&
12.66
&
0
&
1.51
\\
$^{15}\mathrm{O}$
&
12.37
&
0
&
1.58
\\
$^{17}\mathrm{F}$
&
12.37
&
0
&
1.48
\\
\hline
\hline
\end{tabular*}
\vspace{0.1cm}
\protect\caption{\small
The factors
$ Y_{\mathrm{LUM}} $
(see Eq.(\protect\ref{E1012}))
and
the factors
$ \xi_{\mathrm{min}} $
and
$ \xi_{\mathrm{max}} $
which
determine the limits for the values of the total neutrino fluxes
in case B.}
\label{T:FLUXES}
\end{table}

\begin{figure}[t]
\begin{center}
\mbox{\epsfig{file=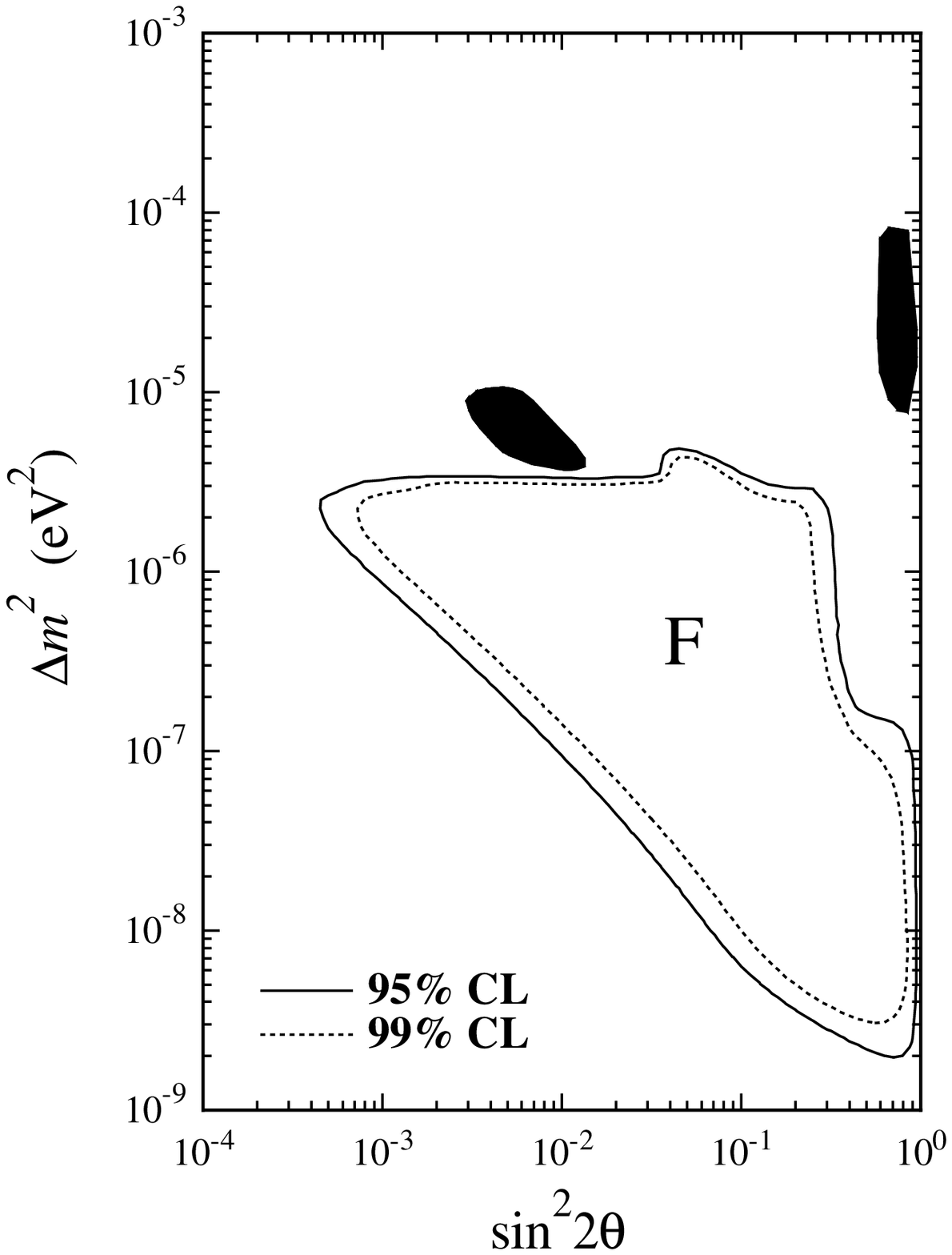,width=\linewidth}}
\end{center}
\protect\caption{\small
Excluded regions in the
$ \sin^2 2\theta $--$ \Delta m^2 $
plane for MSW transitions
due to
$\nu_{e}$--$\nu_{\mu(\tau)}$
mixing in case A.
The region F
is excluded at 95\% CL
within the solid line
and at 95\% CL
within the dotted line.
The allowed regions
found with the BP neutrino fluxes are also shown (shaded areas).
}
\label{F01}
\end{figure}

\begin{figure}[t]
\begin{center}
\mbox{\epsfig{file=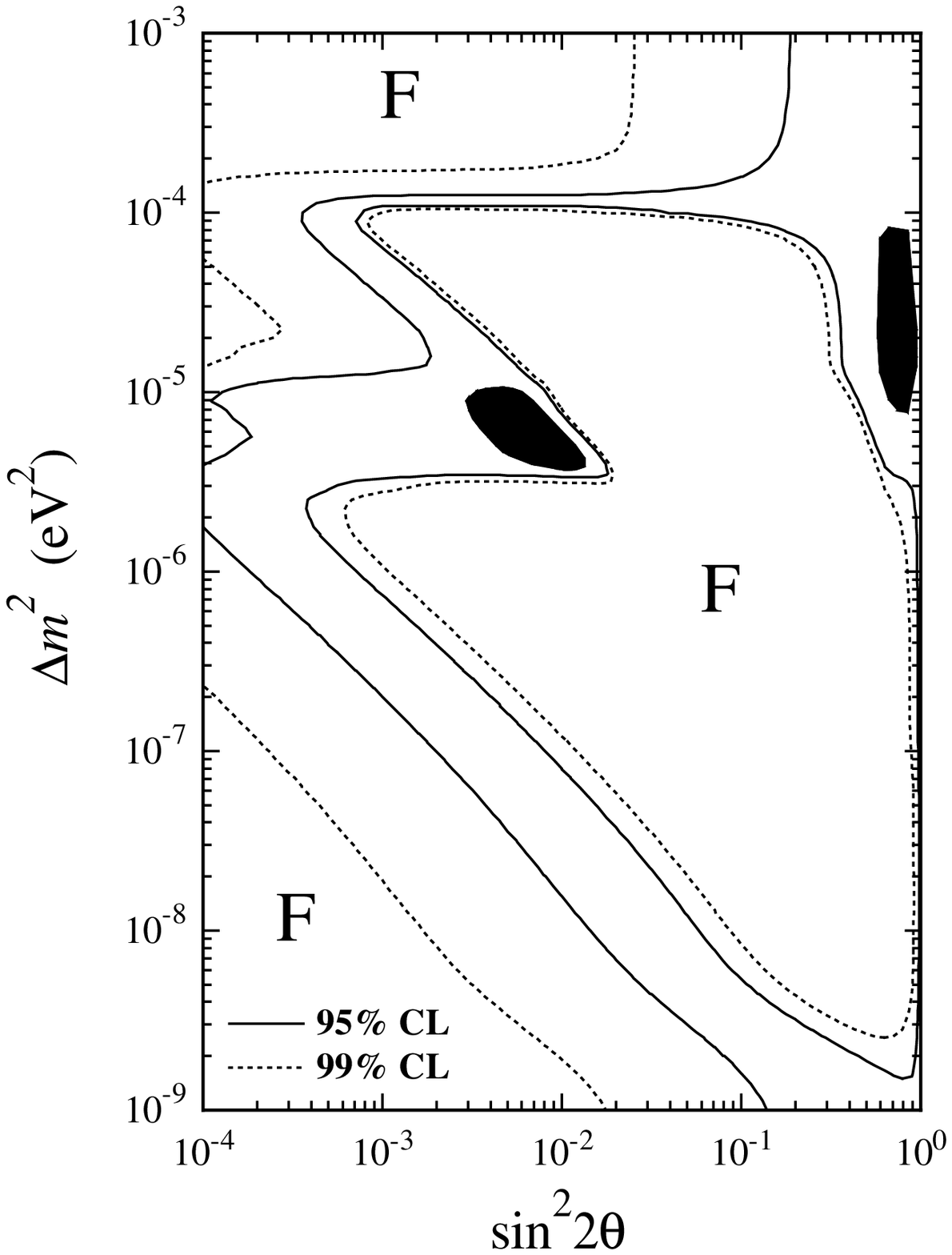,width=\linewidth}}
\end{center}
\protect\caption{\small
Excluded regions in the
$ \sin^2 2\theta $--$ \Delta m^2 $
plane for MSW transitions
due to
$\nu_{e}$--$\nu_{\mu(\tau)}$
mixing in case B.
The regions F
are excluded at 95\% CL
within the corresponding solid line
and at 95\% CL
within the corresponding dotted line.
}
\label{F02}                 
\end{figure}

Our procedure for the analysis of the solar neutrino data
is the following.
At fixed values of the parameters
$ \Delta m^2 $ and $ \sin^2 2\theta $
we calculate the $\chi^2$
for all possible values of the neutrino fluxes.
For each value
of the parameters
$ \Delta m^2 $ and $ \sin^2 2\theta $
and of the neutrino fluxes
we estimate the ``goodness-of-fit''
by calculating the confidence level (CL)
corresponding to the calculated $\chi^2$.
Since we do not determine any parameter,
the number of degrees of freedom
of the $\chi^2$ distribution
is equal to the number of data points
(i.e. four: three neutrino rates and the solar luminosity constraint).
If all the confidence levels
found for a given value of
$ \Delta m^2 $, $ \sin^2 2\theta $
and all possible values of the neutrino fluxes
are smaller than $\alpha$
(we choose $\alpha$ = 0.05, 0.01),
then the corresponding point in the
$ \Delta m^2 $--$ \sin^2 2\theta $
plane is excluded at
$ 100 ( 1 - \alpha ) $\% CL.
In this way we obtain the exclusion plots
presented in Figs.\ref{F01} and \ref{F02}.

For the exclusion plot presented in
Fig.\ref{F01}
the only requirement was that all the total neutrino fluxes
are positive.
Let us call this case A.
However,
it is interesting and instructive to investigate
how the forbidden regions in the
$ \Delta m^2 $--$ \sin^2 2\theta $ plane
change if some limits on the allowed
values of the neutrino fluxes are imposed.
Thus we also considered the following case B:
the different solar neutrino fluxes
are constrained in the interval
$
\xi^{r}_{\mathrm{min}} \, \Phi^{r}(\mathrm{BP})
\le
\Phi^{r}
\le
\xi^{r}_{\mathrm{max}} \, \Phi^{r}(\mathrm{BP})
$,
where
$ \Phi^{r}(\mathrm{BP}) $
are the BP values of the neutrino fluxes
and the factors
$ \xi^{r}_{\mathrm{min}} $
and
$ \xi^{r}_{\mathrm{max}} $
are chosen
in such a way to include the predictions
of the existing solar models
[\ref{L:BAHCALL}--\ref{L:NICE}].
The values of these factors are given in Table~\ref{T:FLUXES}.
We determined the minimum (maximum) values
for the $pp$, $pep$, $^7\mathrm{Be}$ and $\mathrm{He}p$ fluxes
by subtracting (adding) 3 times the range of solar model predictions
to the minimum (maximum) predicted flux.
Since it has been recently suggested
\cite{B:S17}
that
the value of the astrophysical factor
$S_{17}(0)$ could be significantly lower
than that used in SSM calculations,
we let the $^8\mathrm{B}$ flux
to be arbitrarily small.
Since the CNO fluxes
have large uncertainties,
we allow also them
to be arbitrarily small.
We determined the maximum values
of the $^8\mathrm{B}$ and CNO fluxes
by adding 3 times
the $1\sigma$
error of BP
to the BP average value.
Let us emphasize that the limits
on the allowed
values of the neutrino fluxes which we imposed in case B
are rather large.
The excluded regions
of the parameters
$ \Delta m^2 $ and $ \sin^2 2\theta $
in case B are presented in Fig.\ref{F02}.

The excluded region
in case A (Fig.\ref{F01})
has the triangular shape
typical of a strong $\nu_{e}$ suppression.
In this region the value of the flux 
of $pp$ $\nu_{e}$'s on the earth
is strongly suppressed.
This suppression of the flux of $pp$ $\nu_{e}$'s
is in contradiction with the Gallium data.
It cannot be counterbalanced
by a large initial $pp$ flux
$\Phi^{pp}$
because
$\Phi^{pp}$
is limited by the luminosity constraint.
The observed Gallium event rate
cannot be due to a high value of
$ \Phi^{^8\mathrm{B}} $
because the $^8\mathrm{B}$ neutrino flux
is constrained
by the data of the Kamiokande experiment
in which both $\nu_{e}$ and $\nu_{\mu}$ are detected
(notice that in most of the excluded region
the $^8\mathrm{B}$ $\nu_{e}$ flux on the earth
is not suppressed
because the corresponding MSW transition is highly non--adiabatic).
The observed Gallium event rate
cannot either be due to high values of
the other neutrino fluxes
($pep$, $^7\mathrm{Be}$ and CNO)
because the values of these fluxes
are constrained by
the data of the Homestake experiment.

Let us now
consider the excluded regions
in case B
(Fig.\ref{F02}).
A comparison of Fig.\ref{F01} and Fig.\ref{F02}
shows that the region of values of the parameters
$ \Delta m^2 $ and $ \sin^2 2\theta $
that is forbidden
by the existing solar neutrino data
is strongly increased
if we put some limits on the possible values
of the neutrino fluxes.
Fig.\ref{F02}
illustrates the fact
that even assuming rather wide limits
for the values of the solar neutrino fluxes,
the existing solar neutrino data strongly restrict
the region of possible values of the parameters
$ \Delta m^2 $ and $ \sin^2 2\theta $.
The triangular excluded region
for
$ \Delta m^2 \lesssim 5 \times 10^{-6} \, \mathrm{eV}^2 $
is due to a strong suppression of the flux of low energy $pp$ $\nu_{e}$'s.
The excluded region 
with
$ 5 \times 10^{-6} \, \mathrm{eV}^2 \lesssim
\Delta m^2 \lesssim 10^{-4} \, \mathrm{eV}^2 $
is due to a large suppression of the flux of $^8\mathrm{B}$ $\nu_{e}$'s.
In this region,
corresponding to a large suppression of $^8\mathrm{B}$ $\nu_{e}$'s,
it is also impossible to fit the Clorine and Gallium data:
the large flux of $^7\mathrm{Be}$ neutrinos
that is necessary to fit the Homestake data
would give an excessive Gallium event rate.
In the large (left) part of the excluded region
$ \mathrm{P}_{\nu_{e}\to\nu_{e}} = 1 $,
which is excluded at 99\% CL.

In conclusion,
we have presented the results of a model independent analysis
of the existing solar neutrino data
in the case of MSW transitions
between two active neutrino types
($ \nu_{e} $--$ \nu_{\mu} $
or
$ \nu_{e} $--$ \nu_{\tau} $).
We have shown
that in this model independent approach
the existing solar neutrino data allow to exclude
rather large regions of values of the parameters
$ \Delta m^2 $ and $ \sin^2 2\theta $
(expecially when limits on the values
of the total neutrino fluxes are imposed).

\end{document}